\documentclass[aip,reprint]{revtex4-1}

\usepackage{graphicx}
\usepackage[colorlinks,urlcolor=blue,citecolor=blue,linkcolor=blue]{hyperref}
\usepackage{amsmath}
\usepackage{amssymb}
\usepackage{enumitem}

\usepackage[dvipsnames]{xcolor}
\definecolor{goodgreen}{rgb}{0.1,0.5,0}
\definecolor{goodred}{rgb}{0.7,0,0}
 
\newcommand{\red}[1]{{\color{goodred}{#1}}} 

\draft 

\begin{document}


\title{Higher-order topology in plasmonic kagome lattices} 





\author{Matthew Proctor}
\email[]{matthew.proctor12@imperial.ac.uk}
\affiliation{%
Department of Mathematics, Imperial College London, London, SW7 2AZ, UK}

\author{Mar\'{i}a Blanco de Paz}
\affiliation{Donostia International Physics Center, 20018 Donostia-San Sebasti\'an, Spain}

\author{Dario Bercioux}
\affiliation{Donostia International Physics Center, 20018 Donostia-San Sebasti\'an, Spain}
\affiliation{IKERBASQUE, Basque Foundation for Science, Euskadi Plaza, 5, 48009 Bilbao, Spain}

\author{Aitzol Garc\'{i}a-Etxarri}
\affiliation{Donostia International Physics Center, 20018 Donostia-San Sebasti\'an, Spain}
\affiliation{IKERBASQUE, Basque Foundation for Science, Euskadi Plaza, 5, 48009 Bilbao, Spain}

\author{Paloma Arroyo Huidobro}%
\affiliation{%
Instituto de Telecomunica\c c\~oes, Instituto Superior Tecnico-University of Lisbon, Avenida Rovisco Pais 1, Lisboa, 1049‐001 Portugal}%

\date{\today}


\begin{abstract}
We study the topological properties of a kagome plasmonic metasurface, modelled with a coupled dipole method which naturally includes retarded long range interactions. We demonstrate the system supports an obstructed atomic limit phase through the calculation of Wilson loops. Then we characterise the hierarchy of topological boundary modes hosted by the subwavelength array of plasmonic nanoparticles: both one-dimensional edge modes as well as zero-dimensional corner modes. We determine the properties of these modes which robustly confine light at subwavelength scales, calculate the local density of photonic states at edge and corner modes frequencies, and  demonstrate the selective excitation of delocalised corner modes in a topological cavity, through non-zero orbital angular momentum beam excitation.

\end{abstract}


\maketitle 

\section{Introduction}\label{sec:intro}

Two-dimensional (2D) higher-order topological insulators (HOTIs) host protected zero-dimensional (0D) localised corner states, in addition to conventional one-dimensional (1D) edge states.
Following their theoretical proposal in condensed matter~\cite{Benalcazar2017Quantized} and subsequent experimental demonstration in electrical circuits~\cite{Imhof2018topoelectrical}, many of the concepts have been transferred to classical wave systems including photonics~\cite{Kim2020recent}.
Early proposals in photonics focused on realisations of corner modes in evancescently coupled waveguides~\cite{ElHassan2019corner,Noh2018}, which map directly to tight-binding models with nearest neighbour interactions due to the short-range nature of the coupling.
More recent literature has turned to photonic crystals, where the interactions are necessarily long-range~\cite{proctor2020robustness}.
A number of experimental investigations have shown how corner modes can arise in photonic crystals with different crystalline symmetries, such as $C_3$~\cite{smirnova2020room}, $C_4$~\cite{han2020lasing} and $C_6$~\cite{Xie2020higher}.

In particular, the $C_3$-symmetric `breathing kagome' lattice and its hierarchy of topological modes have attracted considerable attention~\cite{ezawa2019higher,Kempkes2019,Ni2017topological,Ni2019observation,ni2017topologicalproc}.
It hosts topological valley edge states, due to the localisation of Berry curvature with opposite sign at opposite valleys~\cite{wong2020gapless,saba2020nature}.
Notably, it also has a topologically-distinct bulk as a result of displaced Wannier centers~\cite{ezawa2019higher} which result in corner modes in finite lattices.
These corner modes have been probed experimentally in coupled waveguides~\cite{ElHassan2019corner} and recently in a photonic crystal~\cite{Li2020}.
There it was shown that by adding ad hoc long-range interactions to a next\red{-}nearest neighbour tight-binding model, an additional set of trivial corner modes appears.

In this Letter, we investigate higher-order topology in a subwavelength dipolar array:
The breathing kagome plasmonic metasurface.
To study the hierarchy of subwavelength topological modes emerging in this system, and the effects of slowly decaying photonic interactions,  we consider a coupled dipole model which naturally incorporates full electromagnetic interactions.
We start by calculating Wilson loops for the system in order to characterise the bulk topology.
To appropriately capture the physical behaviour of the metasurface, we calculate the optical response including retarded interactions.
Next, we reveal the spin angular momentum properties of the valley topological edge states and their directional excitation.
Moving to a finite system, we show how topological corner states emerge and explore their  robustness.
Finally we study methods of exciting these modes,
firstly by probing the local density of states (LDOS) of the corner and edge modes, and then by showing the selective excitation of corner modes.


\section{Methods}\label{sec:methods}

We model the metasurface, a 2D lattice of metallic nanoparticles (NPs), with the coupled dipole method (CDM). 
We consider silver nanorods modelled with the analytical spheroid polarisability~\cite{moroz2009depolarization}, which includes depolarization and radiative effects to appropriately describe the optical response.
The dielectric permittivity given by a Drude model, $\epsilon(\omega)=\epsilon_\infty-\omega_p^2/(\omega+i\gamma)$, with $\omega_p = 8.9$~eV, $\gamma = 38$~meV and $\epsilon_\infty = 5$~\cite{yang2015optical}. 
We only consider the out-of-plane modes here and note that the in-plane modes, which occur at a different frequency due to the spheroidal NPs, have been studied previously in a dipolar breathing kagome lattice~\cite{zhang2020second}, although only nearest and next-nearest neighbor interaction terms are included.
Importantly, interactions between NPs are given by the full dyadic Green's function, which incorporates retarded interactions and explicitly compare to the quasistatic (QS) approximation; showing its importance in describing the physical behaviour of a plasmonic metasurface~\cite{proctor2019exciting}. 
The CDM is also a suitable model for any system of subwavelength dipolar scatterers, such as silicon carbide NPs~\cite{zhang2020second}, quarter-wavelength resonators~\cite{yves2020locally} or cold atoms~\cite{perczel2017photonic}.
More details can be found in the Supplementary Material (SM).

\section{Topology of the bulk modes}

\begin{figure}
    \centering
    \includegraphics{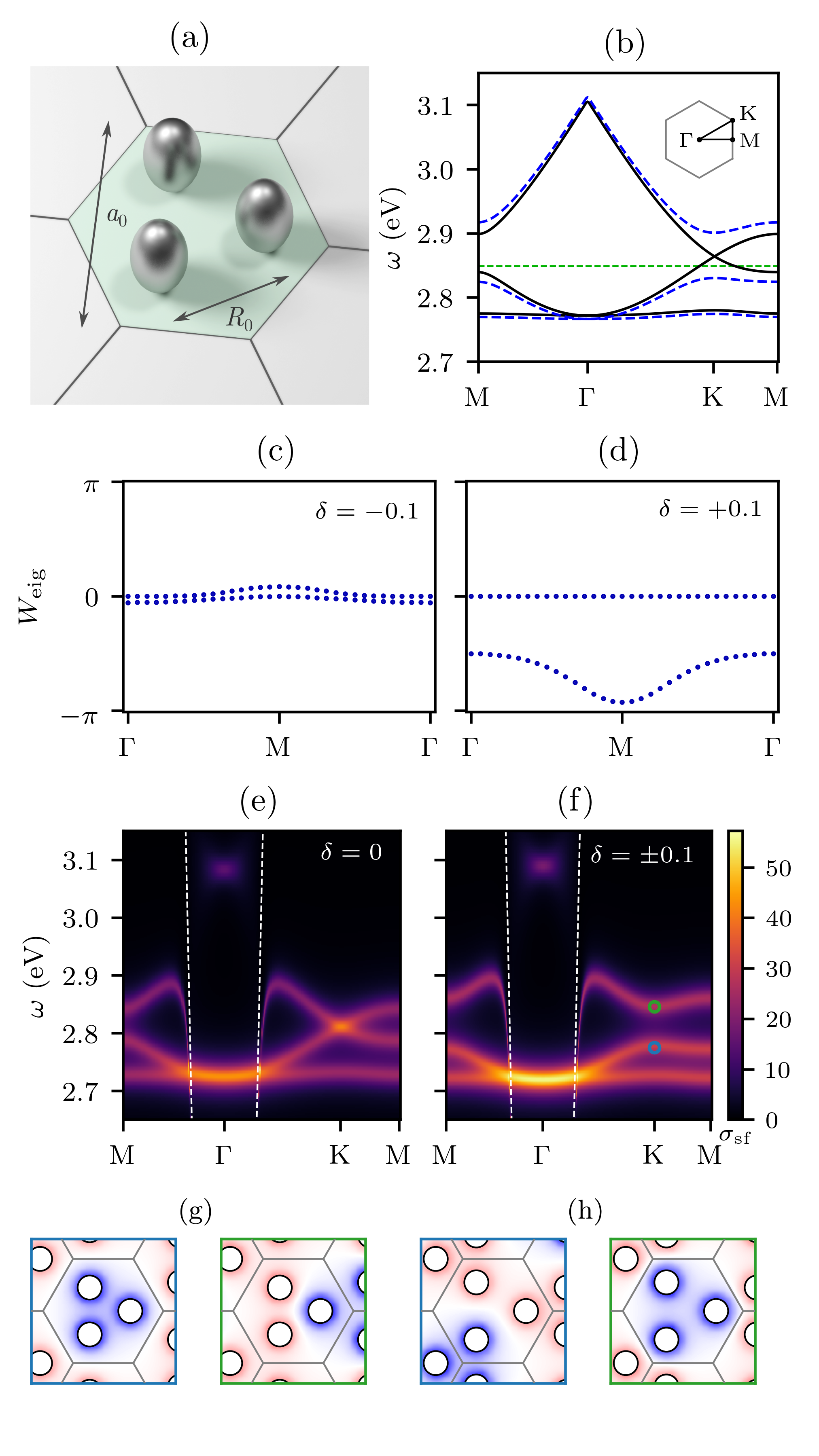}
    \caption{\label{fig:bulk_response}
    Topology of the bulk modes:
    (a) Unit cell. Lattice constant $a_0 = 50\sqrt{3}$~nm and nearest neighbour spacing $R_0 = 25\sqrt{3}$~nm. NPs are silver with radius $r = 10$~nm and height $h = 40$~nm. The lattice is perturbed by contracting or expanded the trimer by a scale factor $\delta$, $R = (1 + \delta)R_0$.
    (b) Band structure, QS approximation: Unperturbed $\delta = 0$ in black and perturbed $\delta = \pm 0.1$ in blue. The contracted and expanded lattice bands are identical. $\omega_\mathrm{lsp}$ shown as the horizontal green line.
    Wilson loop eigenvalues for the (c) Contracted and (d) expanded lattices. 
    The contracted lattice is trivial and the expanded is in a photonic obstructed atomic limit phase.
    Spectral function $\sigma_\mathrm{sf}$ with fully electrodynamical interactions for the (e) unperturbed lattice and (f) perturbed lattices.
    $E_z$ of the eigenmodes at $K$ for the (g) contracted and 
    (h) expanded lattices, for the band below (blue) and above (green) the topological band gap, as indicated as dots in (f). 
    }
\end{figure}

We begin by calculating the spectrum in the QS approximation,
however we go beyond a nearest-neighbour (NN) approximation  and include all neighbours in the lattice sum.
The kagome unit cell is shown in Fig.~\ref{fig:bulk_response}(a).
In Fig.~\ref{fig:bulk_response}(b), we plot the band structure of the unperturbed kagome lattice ($\delta = 0$) in black.
The cusp at $\Gamma$ is unphysical, as we shall show, but is typical of including all neighbours in the QS lattice sum~\cite{proctor2019exciting,zhen2008collective}.
At $K(K')$ we see a Dirac cone between the two highest energy bands, associated with the $C_{3v}$ symmetry of the lattice~\cite{proctor2019manipulating}.
To gap the Dirac cone, we break inversion symmetry, $\sigma_v$, by perturbing the trimer in the unit cell inwards or outwards by a scale factor $\delta$, such that $R = (1 + \delta)R_0$. 
We will refer to these as the `contracted' and `expanded' lattices, respectively.
We plot the bands for these lattices with $\delta = \pm0.1$ as dashed blue lines in Fig.~\ref{fig:bulk_response}(b), noting that both perturbed lattices have identical band structures~\cite{Ni2017topological, wong2020gapless, saba2020nature}.

\red{E}ven though these
lattices share band structures, they can be distinguished by their Wilson loops~\cite{depaz2020tutorial}.
Eigenvalues of the Wilson loop operator describe the evolution of Wannier centers along closed paths in the Brillouin zone and provide information about the band's topological characters~\cite{depaz2019engineering}.
We calculate Wilson loops in the QS approximation for the two connected bands below the topological band gap, for the two lattice configurations.
While retardation can have significant effects on topological properties, we emphasise here that we are in a subwavelength parameter regime~\cite{pocock2018topological,pocock2019bulk}.
For the contracted lattice, Fig.~\ref{fig:bulk_response}(c), the Wilson loops are displaced about $W = 0$. 
However for the expanded lattice, Fig.~\ref{fig:bulk_response}(d), they are displaced about $W = 0$ as well as away from zero.
Lattices with this behaviour are in an obstructed atomic limit phase~\cite{depaz2019engineering}.
We note that unlike previous works, our calculations of Wilson loops go beyond a NN approximation~\cite{Ni2019observation}. 

\red{T}o capture the optical response of this fully electrodynamical system, we use the retarded Green's function and the radiative polarizability to calculate the spectral function $\sigma_\mathrm{sf}$.
We plot this for the unperturbed, Fig.~\ref{fig:bulk_response}(e), and  perturbed lattices, Fig.~\ref{fig:bulk_response}(f).
Firstly, $\sigma_\mathrm{sf}$ shows the broadening 
and slight redshifting of the modes due to the radiative effects. 
Starting at $\Gamma$ the highest energy mode has a monopolar character and the two lowest energy modes are degenerate dipolar modes.
At the light line, a strong polariton-like splitting occurs in the highest energy monopolar band due to the coupling to free photons; similar to effects in 1D~\cite{koenderink2006complex} and other 2D plasmonic lattices~\cite{fernique2019plasmons, proctor2020near,Kolkowski2020lattice}.
Notably, the fully electrodynamical interactions remove the  cusp at $\Gamma$ observed in the QS approximation.

The system also has a band inversion at $K/K'$ between the contracted and expanded phases~\cite{saba2020nature}.
We plot the out of plane electric field $E_z$ for the two bands involved in the Dirac cone gapping, for contracted, Fig.~\ref{fig:bulk_response}(g) , and expanded, Fig.~\ref{fig:bulk_response}(h), phases.
In the contracted phase, the highest energy band is monopolar and the next band is dipolar, while the ordering is flipped in the expanded phase.
Finally, we note that the overall ordering here is opposite to the dielectric system~\cite{wong2020gapless}, and the flat band is the lower frequency band. 
This is due to the negative and dispersive permittivity of metallic NPs and is not a result of the dipolar nature of the model, as claimed elsewhere~\cite{zhang2020second}. 

\section{Topological valley edge states}

\begin{figure}
    \centering
    \includegraphics{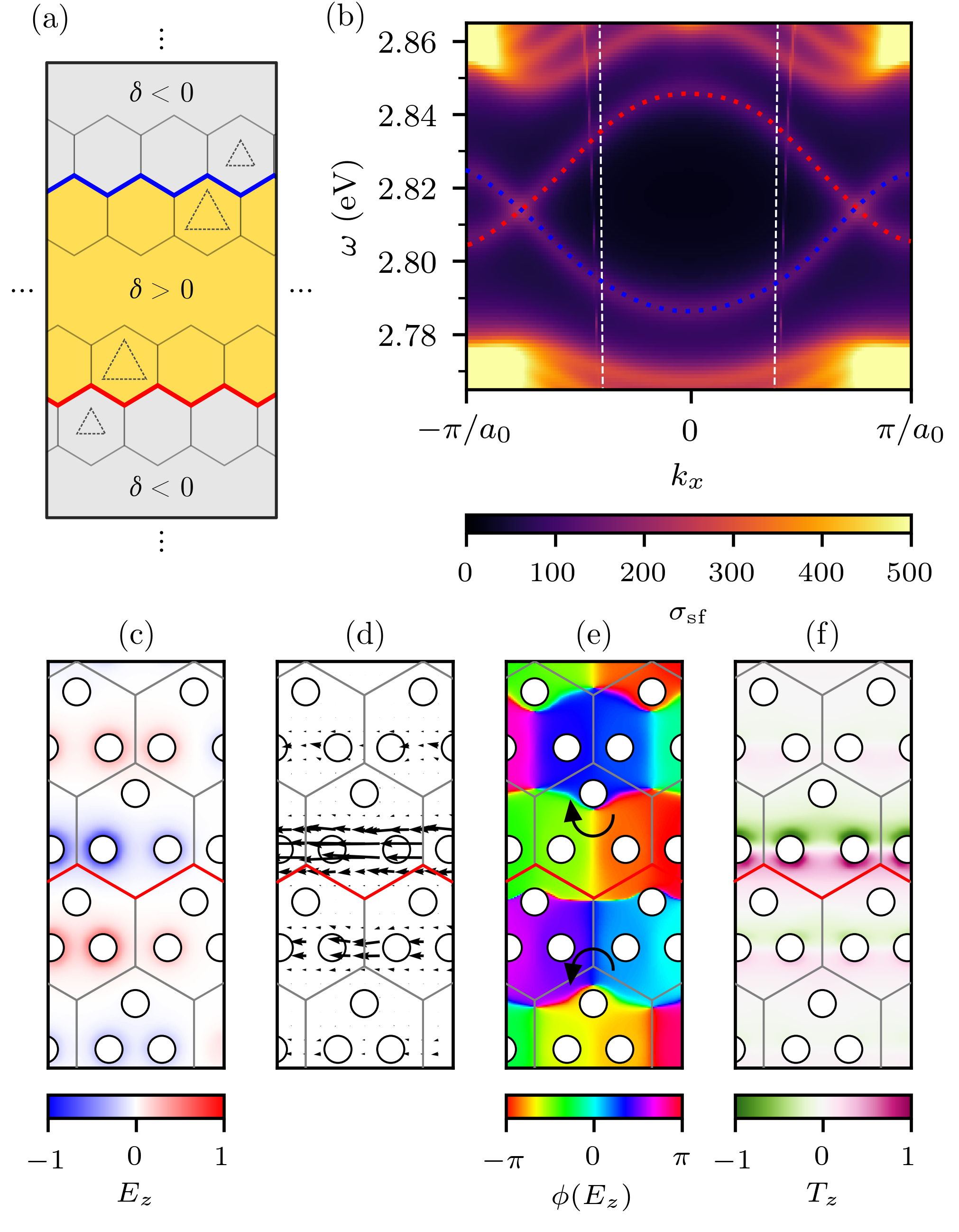}
    \caption{\label{fig:edge_modes}Valley topological edge modes, (a) 
    Ribbon ($20\times1$ unit cells) with an interface between expanded $\delta = +0.1$ (grey) and contracted $\delta = -0.1$ (yellow) regions. (b) Spectral function $\sigma_\mathrm{sf}$, with edge states for the expanded/contracted (blue) and contracted/expanded (red) interfaces. Losses are $\gamma = 10$~meV to increase the visibility of the edge states. 
    Edge state eigenmodes at $k_x = +\frac{1}{2}\frac{\pi}{a_0}$, (c) Out-of-plane electric field $E_z$, (d) In–plane time-averaged Poynting vector $\langle\mathbf{S}\rangle$, (e) out-of-plane electric field phase $\phi(E_z)$ and (f) out-of-plane spin angular momentum $T_z$.}
\end{figure}

We now move onto investigate the topological valley edge states which arise
at the interface between contracted and expanded regions. 
These are guaranteed to occur via the bulk boundary correspondence, provided the perturbation between the two regions $\delta$ is small enough and preserves the localisation of Berry curvature at the $K/K'$ valleys~\cite{qian2018topology, makwana2018geometrically}.
Propagation along these interfaces is protected against disorder which does not mix the valleys due to intervalley separation~\cite{orazbayev2019quantitative}.
To model these edge states, 
we set up a
ribbon with interfaces between the two phases as shown in Fig.~\ref{fig:edge_modes}(a), with $\delta = \pm0.1$ | see SM.
The ribbon spectral function, Fig.~\ref{fig:edge_modes}(b), shows how
within the band gap 
there are two edge states; the red band corresponds to the expanded over contracted (E/C) interface and the blue band to the contracted over expanded (C/E) one. 

To calculate the edge mode profiles we linearise the Green's function, see SM. In Fig.~\ref{fig:edge_modes}(c) we plot the out-of-plane electric field $E_z$ for the  E/C interface, which shows how the edge state is strongly confined to the interface. 
Then we show the in-plane time averaged Poynting vector $\langle\mathbf{S}\rangle = \frac{1}{2}\mathrm{Re}(\mathbf{E}^*\times\mathbf{H})$ (d), out-of-plane electric field phase $\phi(E_z)$ (e) and spin angular momentum $\mathbf{T} = \frac{1}{2}\mathrm{Im}(\mathbf{E}^*\times\mathbf{E} + \mathbf{H}^*\times\mathbf{H})$~\cite{bliokh2017optical} (f), which encodes the degree of elliptical polarisation of in-plane magnetic fields and is out of the plane $\mathbf{T} = T_z\hat{\mathbf{z}}$ for $z = 0$.


From the Poynting vector $\langle\mathbf{S}\rangle$, we see how the energy flow is confined to the interface, with direction given by the group velocity $v_g < 0$ of the mode at $k_x>0$, 
while $\phi(E_z)$ and $T_z$ determine the propagation direction of modes upon excitation~\cite{proctor2020near}. 
For example, beams with non-zero orbital angular momentum will excite a directional mode if the phase vortex of the beam matches the phase vortex in $\phi(E_z)$ of the edge eigenmode: this occurs at the centre of an expanded unit cell next to the interface and has been shown in a similar photonic crystal~\cite{deng2019vortex,gong2020topological}. 
Additionally, 
if a circularly-polarised in-plane magnetic dipole source is placed at the centre of an expanded unit cell, it will excite a mode along a direction determined by the source polarisation: right(left)-propagating for right(left) circular polarisation.
However, at the edge of the unit cell the local handedness of polarisation changes sign and sources will couple to the mode propagating in the opposite direction.
We show examples of directional propagation in the SM.

\section{Higher-order topological modes}\label{sec:emergence_corner}

Before studying the emergence of corner modes, let us briefly review the concept of chiral symmetry.
In the limit of short range NN interactions, lattices with an even number of elements in the unit cell are chirally symmetric~\cite{asboth2016short}. In condensed-matter systems, chiral symmetry pins corner modes to a specific frequency in \mbox{HOTIs} \cite{Peterson2020fractional}. 
This imposes a subtle distinction between HOTIs and photonic systems with long-range couplings,
where chiral symmetry is necessarily broken due to interactions between sites in the same sublattice.
By continuously reducing the interaction range we have previously shown that retarded systems are deformable to ones with chiral symmetry~\cite{proctor2020robustness}: Meaning they possess an approximate chiral symmmetry.
We refer to corner modes in these systems as `higher-order topological modes' (HOTMs).

\begin{figure}
    \centering
    \includegraphics{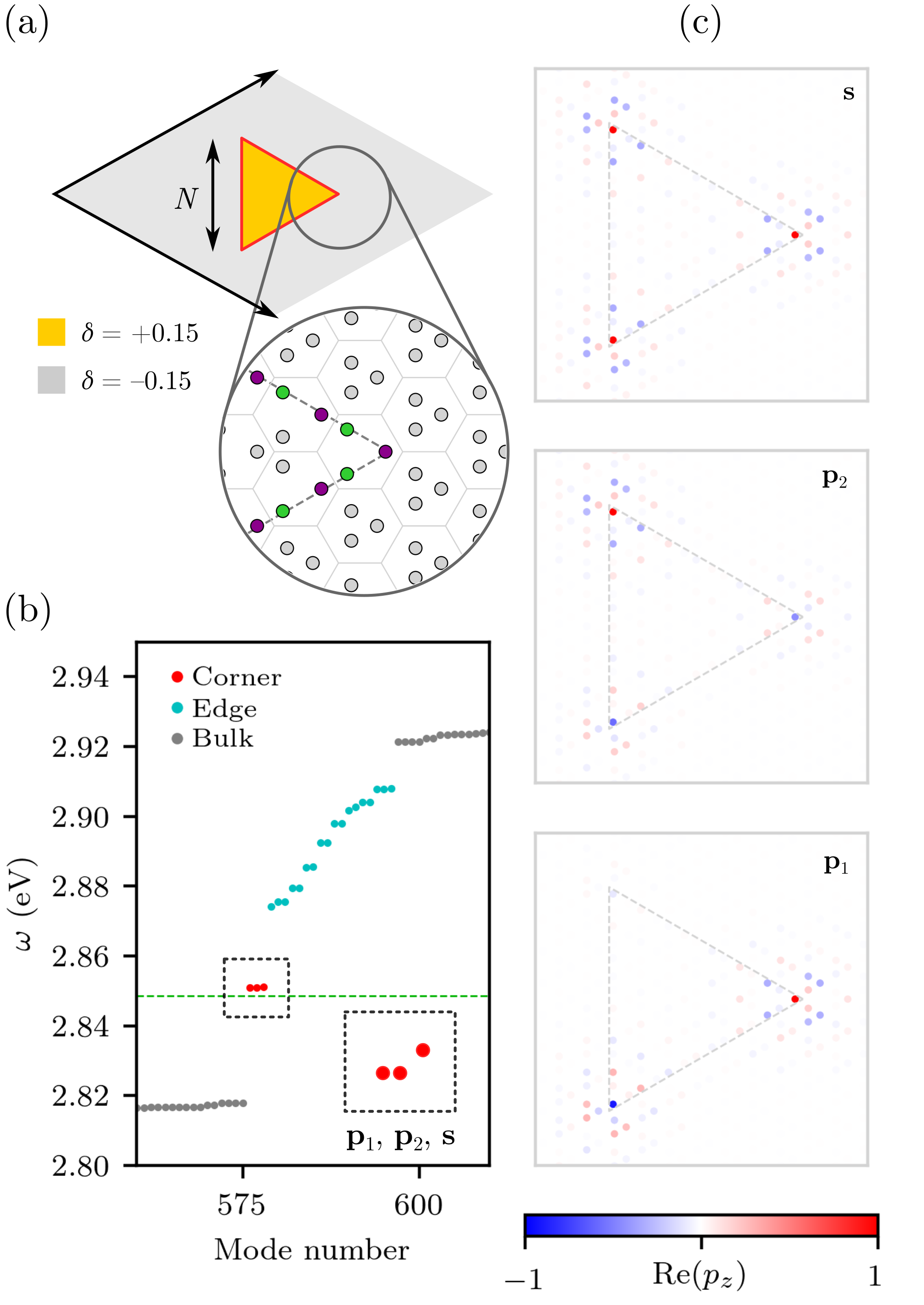}
    \caption{\label{fig:corner_eigenmodes}Higher-Order Topological Modes,
    (a) The topological particle (yellow) consists of unit cells in the expanded phase, $\delta = +0.15$ and is surrounded the contracted phase, $\delta = -0.15$. The side of the particle is $N = 7$ unit cells. Inset: A close of the NP arrangements at the corner.
    (b) Spectrum of topological particle modes against mode number, with corner (red), edge (cyan) and bulk (grey) modes. 
    (c) Dipole moments $p_z$ of the three corner modes. Mode $\mathbf{s}$ (highest energy) has a monopolar character and modes $\mathbf{p}_{1,2}$ (lowest energy) are dipolar.}
\end{figure}

The breathing kagome lattice has an odd number of elements in the unit cell, meaning it is incompatible with chiral symmetry. 
On the other hand, a `generalised chiral symmetry' has been proposed for this system~\cite{Ni2017topological,Ni2019observation}.
This acts similarly to chiral symmetry in that, in the limit of short-range interactions, it pins the corner modes at a mid-gap energy (or the localised surface plasmon frequency, $\omega = \omega_\mathrm{lsp} = 2.845$~eV for these plasmonic NPs) and forces the excitation on a single sublattice in real space.
With long-range interactions, the breathing kagome lattice still has an approximate generalised chiral symmetry~\cite{Li2020} as a result of critical interactions between the same sublattice being negligible compared to interactions between different sublattices.
As such, long-range interactions only slightly shift the frequency of the corner modes away from mid-gap energy, $\omega = \omega_\mathrm{lsp}$. 

To investigate corner modes 
we set up a periodic supercell containing a region in the expanded phase (the `topological particle') surrounded by the contracted phase, Fig.~\ref{fig:corner_eigenmodes}(a).
The finite nature of the topological particle means there will be a discrete set of modes which arise on the interface~\cite{siroki2017topological}.


\begin{figure}
    \centering
    \includegraphics{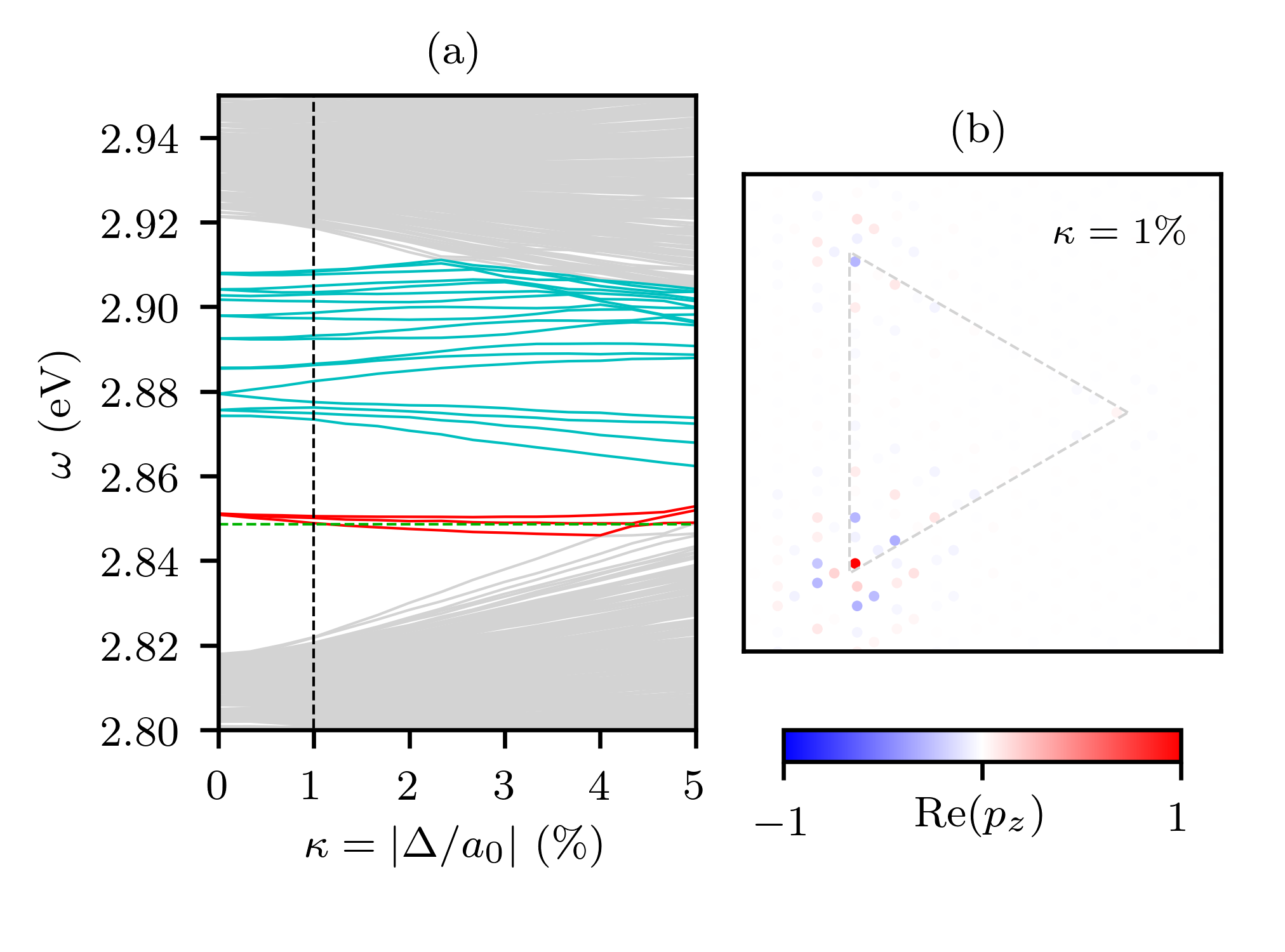}
    \caption{\label{fig:disorder}Effect of random positional disorder,
    (a) Spectrum for increasing disorder. The HOTMs are lost to the bulk at approximately $\kappa = 4\%$.
    (b) Dipole moments $p_z$ of the three corner modes for $\kappa = 1\%$ (black dotted line in (a)). Excitations on multiple corners are still possible even with spatial disorder owing to retardation.}
\end{figure}

Since HOTMs are reliant on the displaced Wannier centres~\cite{ezawa2019higher}, rather than the localisation of Berry curvature as the edge modes are,
we are free to choose the perturbation $\delta > 0$.
This controls the size of the band gap as well as the localisation of modes to the corners, see SM. 
Again we linearise the Green's function to calculate the 
spectrum of the topological particle.
Fig.~\ref{fig:corner_eigenmodes}(b) shows the spectrum for $\delta = 0.15$ with corner modes (red) and edge modes (cyan) appearing within gapped bulk modes (grey).
In Fig.~\ref{fig:corner_eigenmodes}(c) we plot the real part of the out-of-plane dipole moments $\mathrm{Re}(p_z)$ of the three corner eigenmodes. 
For a system with short-range interactions, the eigenmodes are triply degenerate~\cite{ElHassan2019corner}, whereas here the long-range interactions split the degeneracy of the eigenmodes into $(1+2)$ as dictated by symmetry of the topological particle. 
Mode $\mathbf{s}$ is a monopolar-like mode, with the NPs at the three corners oscillating in phase, whilst modes $\mathbf{p}_1$ and $\mathbf{p}_2$ are dipolar-like modes. 
Although the effects of long-range interactions on corner modes in the breathing kagome lattice have been studied previously in photonic crystals~\cite{Li2020}, here we start with a model that includes fully retarded photonic interactions; such that there is coupling between all sites in the lattice.



The HOTMs
are protected by the approximate generalised chiral symmetry and crystalline symmetry of the particle.
We demonstrate the robustness by plotting the mode spectrum for increasing random positional disorder in Fig.~\ref{fig:disorder}(a). 
To apply the disorder, we shift each NP randomly in the $xy$-plane up to a percentage $\kappa = |\boldsymbol\kappa|$ of the lattice constant $a_0$, $\mathbf{r} \rightarrow \mathbf{r} + \boldsymbol\kappa a_0$.
Since positional disorder breaks $C_3$ and mirror symmetries across the topological particle, the $(1+2)$-degeneracy is broken. 
Despite this the HOTMs are robust up to $\kappa \approx 4\%$ where they eventually merge with the bulk.
In Fig.~\ref{fig:disorder}(b) we plot a HOTM
for $\kappa = 1\%$, indicated by the vertical black dotted line in panel (a). 
The slowly-decaying nature of retarded interactions allows the excitation of modes delocalised over the cavity even if the crystalline symmetry is broken, provided the size of the cavity is not too large.
In comparison, in the QS approximation (not shown) the modes become localised to single corners.
This highlights the limitations of tight-binding
models when studying nanophotonic systems~\cite{ElHassan2019corner}, and explains how cavity delocalised corner modes can be observed in experimental setups~\cite{smirnova2020room}, despite disorder due to fabrication errors. 
Finally, we note that another indicator of higher-order topology is a fractional corner anomaly~\cite{Peterson2020fractional}.
The breathing kagome plasmonic metasurface presents this behaviour, as we show in the SM, which justifies referring to localised corner states as examples of higher-order topology.

\section{Excitation of Topological Modes}\label{sec:selective_excitation}

\begin{figure}
    \centering
    \includegraphics{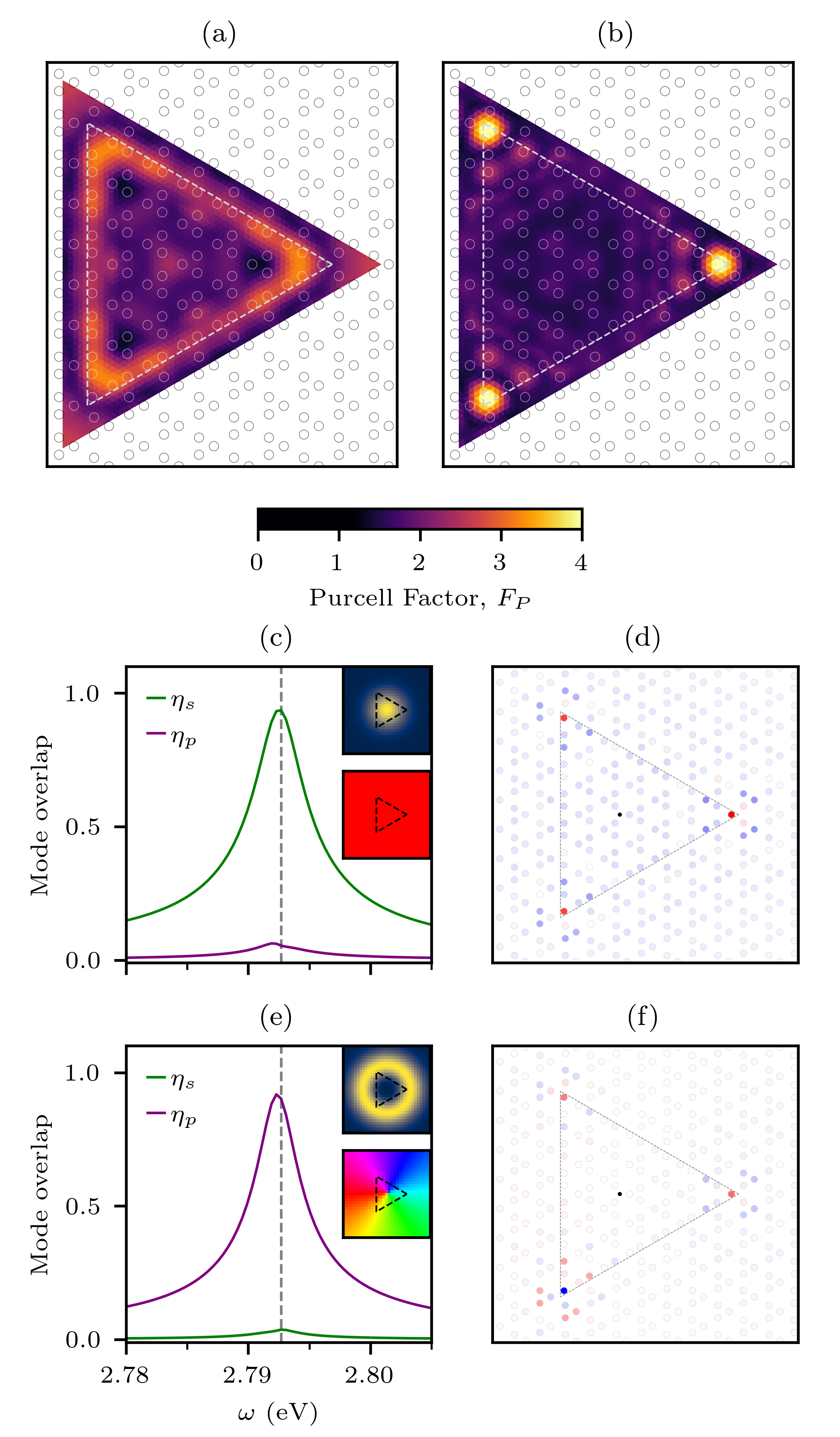}
    \caption{\label{fig:selective_excitation}
    Excitation of topological boundary modes. 
    Purcell factor for out-of-plane electric dipole source at $z = 60$~nm for,
    (a) Edge state frequency $\omega = 2.8445$~eV and
    (b) HOTM frequency $\omega = 2.7925$~eV.
    Selective excitation with Laguerre-Gaussian beams.
    Amplitudes and phases of the beams are shown in the insets.
    Overlap $\eta$ of excited dipole moments with eigenmode dipole moments for (c) $LG_{00}$ and (e) $LG_{01}$.
    Excited dipole moments $p_z$ for $\omega = 2.7925$~eV for (d) $LG_{00}$ (monopolar mode) and (f) $LG_{01}$ (dipolar mode).}
\end{figure}

Following the understanding of how topological boundary modes emerge, we now investigate methods of exciting them.
We consider the coupling of a near-field source to the edge and corner modes of the topological particle.
We characterise this through calculations of the partial LDOS for a vertical source, $\rho_z(\omega)$, which captures the available photonic states which the source can couple to~\cite{novoty2012nano, mignuzzi2019nanoscale}.
From this we calculate the Purcell factor,
\begin{align}
    F_P&=\frac{\Gamma}{\Gamma_0}=\frac{\rho_z}{\rho_0}= 1 + \frac{3\epsilon_0}{2|\boldsymbol\mu|^2}\frac{1}{k^3}\mathrm{Im}[\boldsymbol\mu^*\cdot\mathbf{E}_s(\mathbf{r}_0)],
\end{align}
with the decay rate enhancement $\Gamma/\Gamma_0$ and LDOS enhancement $\rho_z/\rho_0$ with respect to vacuum.
We consider a vertical dipole source, with dipole moment $\boldsymbol\mu = (0, 0, 1)^T$, and calculate the electric field scattered by the metasurface at the source position, $\mathbf{E}_s(\mathbf{r}_0)$.
We take $\gamma = 1$~meV to make the modes more visible. 
For higher losses, the modes will become broader but will still be visible provided they are well separated from the bulk modes.

We plot $F_P$ for a frequency within the edge states band in Fig.~\ref{fig:selective_excitation}(a).
As expected the $F_P$ is highest at the edges of the topological particle and we note that peaks in $F_P$ are not restricted to a single sublattice.
On the other hand, in Fig.~\ref{fig:selective_excitation}(b) we show how when exciting at the HOTM frequency $F_P$ peaks at the corners and on only one sublattice, due to the generalised chiral symmetry respected by the corner modes.
As we showed earlier, another consequence of this symmetry is the HOTM frequency being close to $\omega_\mathrm{lsp}$.
Therefore the increase in $F_P$ over the NP directly at the corner is a combination of coupling both to a single NP LSP as well as to a HOTM.
Critically, the coupling to the HOTM allows for the excitation of delocalized modes across the topological particle: A point source at one corner couples to the other two corners through the HOTM (see  SM).
This phenomenon could be exploited to generate topological long range effective interactions between quantum emitters~\cite{barik2018topological}.
We will now show how to selectively excite HOTMs from the 
far-field using Laguerre-Gaussian ($LG_{pl}$) beams.
These are higher-order laser modes which can have non-zero orbital angular momentum due to optical vortices in the phase of the electric field~\cite{novoty2012nano}.
Laguerre-Gaussian beams are parameterized by an azimuthal index $l$, which controls the optical vortices, and a radial index $p$, which controls the nodes of the beam.
In the following, we will use $LG_{00}$ (corresponding to a conventional Gaussian beam) and $LG_{01}$.
The electric field intensity and phase of these is shown in the insets in Fig.~\ref{fig:selective_excitation}(c) and~\ref{fig:selective_excitation}(e).
We excite with a beam with waist $w_0 = 600$~nm, such that the excitation extends across the topological particle and vary the frequency over the HOTM frequency.
Additionally, we excite the system slightly off-normal incidence, $\theta = 1^\circ$, in order to couple to out-of-plane modes (with the electric field polarised in $xz$).

To characterise the selective excitation, we calculate the overlap $\eta$ of the excited dipole moments $\mathbf{q}_\mathrm{exc}$ across the system with the monopolar and dipolar eigenmodes calculated earlier, and normalised to the norm of excited dipole moments,
\begin{align}
    \eta_s &= \frac{1}{|\mathbf{q}_\mathrm{exc}|}\langle\mathbf{q}_\mathrm{exc}|\mathbf{s}\rangle\\
    \eta_p &= \frac{1}{|\mathbf{q}_\mathrm{exc}|}\frac{1}{\sqrt{2}}\left(\langle\mathbf{q}_\mathrm{exc}|\mathbf{p}_1\rangle + \langle\mathbf{q}_\mathrm{exc}|\mathbf{p}_2\rangle\right).
\end{align}
In Fig.~\ref{fig:selective_excitation}(c), we see that the $LG_{00}$ beam predominantly excites a monopole (green line), compared to the dipole (purple line). 
This is due to the beam having OAM $l=0$ and exciting each corner of the particle in phase.
Next in panel~\ref{fig:selective_excitation}(e) we show that the $LG_{01}$ beam preferentially excites the dipolar mode, which is due to the OAM of the beam $l=1$.
The phase vortex of the beam at the origin of the system results in a rotating phase pattern which is no longer azimuthally symmetric across the topological particle.
To confirm the monopolar and dipolar selective excitation, we excite the system at fixed frequency
and plot the dipole moments in real space; in Fig.~\ref{fig:selective_excitation}(d) for $LG_{00}$ and Fig.~\ref{fig:selective_excitation}(f) for $LG_{01}$.
We note that any beam with a phase pattern which breaks the $C_3$ symmetry of the system will allow for dipolar modes to be excited. This means, for example, at large angles of incidence one can excite a dipolar mode with an $LG_{00}$ beam.




\section{Conclusion}\label{sec:conclusion}

In summary, we have investigated crystalline-symmetry dependent higher-order topological corner modes in a plasmonic metasurface of metal nanoparticles arranged in a kagome lattice.
The system hosts a hierachy of topological modes, which we characterize appropriately with long-range interactions.
With Wilson loops we identify an obstructed atomic limit bulk phase.
We additionally calculate the optical response of the bulk with retardation to appropriately capture the physical response of the system.
The interfaces between topologically distinct regions of the lattice host topological valley edge modes and we reveal the spin angular momentum mechanism for exciting directional modes.
We complete the hierarchy with zero-dimensional localised corner modes, which are robust to random spatial perturbations due to their topological origin from the bulk.
Motivated by experiment, we probe the modes of the system in both the near and far-field.
We calculate the Purcell factor for a point source above the metasurface to demonstrate optimal positions for exciting edge and corner modes.
Finally, we show how HOTMs can be selectively excited given the orbital angular momentum of a beam. 
Both edge and corner modes in the plasmonic metasurface could be exploited for mediating interactions between quantum emitters, with the benefits of topological protection~\cite{barik2018topological}.



\section*{Supplementary Material}

See Supplementary Material for: (1) Details on coupled dipole method, including form of the polarizability, and Green's function linearization, (2) Berry curvature and directional propagation of edge states, (3) Corner modes localisation and fractional corner anomaly.

\begin{acknowledgments}
M.P. and P.A.H. acknowledge funding from the Leverhulme Trust. 
P.A.H. further acknowledges funding from Funda\c c\~ao para a Ci\^encia e a Tecnologia and Instituto de Telecomunica\c c\~oes under project UID/50008/2020 and the CEEC Individual program with reference CEECIND/03866/2017.
D.B. and A.G.E. acknowledge funding by the Spanish Ministerio de Ciencia, Innovaci\'{o}n y Universidades through Projects Nos. FIS2017-82804-P  and PID2019-109905GA-C22  respectively. D. B. acknowledges funding from the Transnational Common Laboratory.
A.G.E received funding from the Gipuzkoako Foru Aldundia OF23/2019 (ES) project and by Eusko Jaurlaritza grant numbers IT1164-19 and KK-2019/00101.
$Quantum-ChemPhys$.
\end{acknowledgments}

\section*{Data Availability}

The data that support the findings of this study are available from the corresponding author upon reasonable request.

\clearpage
\bibliography{main}

\end{document}



\title{Supplemental Material: Higher-order topology in plasmonic kagome lattices} 

\author{Matthew Proctor}
\email[]{matthew.proctor12@imperial.ac.uk}
\affiliation{%
Department of Mathematics, Imperial College London, London, SW7 2AZ, UK}

\author{Mar\'{i}a Blanco de Paz}
\affiliation{Donostia International Physics Center, 20018 Donostia-San Sebasti\'an, Spain}

\author{Dario Bercioux}
\affiliation{Donostia International Physics Center, 20018 Donostia-San Sebasti\'an, Spain}
\affiliation{IKERBASQUE, Basque Foundation for Science, Euskadi Plaza, 5, 48009 Bilbao, Spain}

\author{Aitzol Garc\'{i}a-Etxarri}
\affiliation{Donostia International Physics Center, 20018 Donostia-San Sebasti\'an, Spain}
\affiliation{IKERBASQUE, Basque Foundation for Science, Euskadi Plaza, 5, 48009 Bilbao, Spain}

\author{Paloma Arroyo Huidobro}%
\affiliation{%
Instituto de Telecomunica\c c\~oes, Instituto Superior Tecnico-University of Lisbon, Avenida Rovisco Pais 1, Lisboa, 1049‐001 Portugal}%

\date{\today}
\maketitle 

\section{Coupled Dipole Method}

\subsection{Formalism}

The metasurface is a 2D lattice of metallic nanoparticles (NPs), which we model using the coupled dipole method (CDM). Each NP can be treated as a point dipole if it is small enough compared to the wavelength (such that the dipolar mode dominates) and if a system of NPs is sufficiently spaced, such that radius $r$ and nearest neighbour spacing $R$ satisfy $R > 3r$. The interactions between NPs are given by the dyadic Green's function,
\begin{align}\label{eqn:dyadic_gf}
    \hat{\textbf{G}}(\textbf{d}, \omega) &= \frac{e^{ikd}}{d} \biggl[
    \biggl(
    1 + \frac{i}{kd} - \frac{1}{k^2d^2}
    \biggr)\hat{\textbf{I}}\nonumber\\
    &- 
    \biggl(
    1 + \frac{3i}{kd} - \frac{3}{k^2d^2}
    \biggr)\textbf{n}\otimes\textbf{n}
    \biggr],
\end{align}
with $d = |\mathbf{d}|$, $\mathbf{n} = \mathbf{d}/d$ the separation  and $k = \sqrt{\epsilon_m}k_0$ the wavenumber (in the following we assume the system is in vacuum such that $\epsilon_m = 1$).
For multiple NPs, the self-consistent coupled dipole equation relates the dipole moment on one NP, $\mathbf{p}_i$, to an incident field, $\mathbf{E}_\mathrm{inc}$, as well as to the dipole moments on other NPs, $\mathbf{p}_j$, as 
\begin{align}
    \frac{1}{\alpha(\omega)}\mathbf{p}_i = \mathbf{E}_\mathrm{inc} + \sum_{i\neq j}\hat{\mathbf{G}}(\mathbf{d}_{ij},\omega)\cdot\mathbf{p}_j,
\end{align}
where the sum runs over all NPs, excluding self-interactions.

The polarisability $\alpha(\omega)$ encodes the optical response of a single NP.
In the following we consider silver nanorods, with dielectric permittivity given by a Drude model, $\epsilon(\omega)=\epsilon_\infty-\omega_p^2/(\omega+i\gamma)$, with $\omega_p = 8.9$~eV, $\gamma = 38$~meV and $\epsilon_\infty = 5$~\cite{yang2015optical}. 
We model the nanorods with the analytical polarisability of a spheroidal NP~\cite{moroz2009depolarization}, incorporating depolarisation and radiative effects.
For completeness, the polarisability $\alpha(\omega)$ is given by,
\begin{align}
    \alpha(\omega) = \frac{\alpha_s(\omega)}{1 - \dfrac{k^2}{l_E}D\alpha_s(\omega) - i\dfrac{2k^3}{3}\alpha_s},
\end{align}
with the static polarisability $\alpha_s$ as,
\begin{align}
  \alpha_s(\omega) = \frac{V}{4\pi}\frac{\epsilon - 1}{1 + L(\epsilon - 1)}.
\end{align}
$D$ and $L$ are dynamic and static geometrical factors, $l_E$ is the spheroid axis half-length and $V$ is the volume of the spheroid. (For a sphere, $D=1$, $L = 1/3$ and $l_E = r$). 
We only consider the out-of-plane modes here and note that the in-plane modes, which occur at a different frequency,  have been studied previously in a dipolar breathing kagome lattice, although in a (next) nearest neighbour
~\cite{zhang2020second}. 
Importantly here we will model the system using the full dyadic Green's function, incorporating retarded interactions across the whole system and explicitly compare to the QS approximation, showing its importance in describing the physical behaviour of a plasmonic metasurface~\cite{proctor2019exciting}. The CDM is also a suitable model for  any system  of subwavelength dipolar scatterers, so many of the conclusions here can be extrapolated to systems such as silicon carbide NPs~\cite{zhang2020second}, quarter-wavelength resonators~\cite{yves2020locally} or even lattices of cold atoms~\cite{perczel2017photonic}.

For a periodic system, we express the  the dipole moments in terms of Bloch functions, $\mathbf{p}_{nm} = \mathbf{p}e^{i\mathbf{k}_\parallel\cdot\mathbf{R}_{nm}}$; with Bloch wavevector $\mathbf{k}_\parallel$ and lattice sites $\mathbf{R}_{nm} = n\mathbf{a}_1 + m\mathbf{a}_2$, given the lattice vectors $\mathbf{a}_{1,2}$.
Following this we can then set up the following system of equations,
\begin{align}\label{eqn:system}
    \left(\frac{1}{\alpha(\omega)}\mathbb{I} - \hat{\mathbf{H}}(\mathbf{k}_\parallel, \omega)\right)\cdot\mathbf{p} = \mathbf{E}_\mathrm{inc}.
\end{align}
The elements of the interaction matrix $\hat{\textbf{H}}(\textbf{k}_\parallel, \omega)$ contain lattice sums,
\begin{align}\label{eqn:interaction_matrix}
    H_{ij}=
    \begin{cases}
    \sum\limits_{n,m} \hat{\textbf{G}}(\textbf{d}_{ij} + \textbf{R}_{nm}, \omega) \hspace{2px} e^{-i\textbf{k}_\parallel\cdot\textbf{R}_{nm}} & i \neq j\\
    \sum\limits_{{\substack{n\neq0\\m\neq0}}} \hat{\textbf{G}}(\textbf{R}_{nm}, \omega) \hspace{2px} e^{-i\textbf{k}_\parallel\cdot\textbf{R}_{nm}} & i = j
    \end{cases},
\end{align}
which run over the lattice sites.
From Eq.~\eqref{eqn:interaction_matrix}, we can calculate observable optical properties of the metasurface, such as the extinction cross section.
Alternatively, we can also calculate the spectral function $\sigma_\mathrm{sf}$ by letting $\mathbf{E}_\mathrm{inc} = 0$ from the effective polarisability $\alpha_\mathrm{eff} = \sum_n{1/\lambda_n}$, with eigenvalues $\lambda_n$ of the system of equations in Eq.~\eqref{eqn:system}, as $ \sigma_\mathrm{sf} = k\mathrm{Im}\left(\alpha_\mathrm{eff}\right)$.

\subsection{Linearised Green's function}

The coupled dipole equation, Eq.~(3) of the main text, can be used to calculate the modes of the system in various ways. The spectral function can be obtained by varying $\mathbf{k}_\parallel$ and $\omega$, and calculating eigenvalues at each point as explained in the main text.
For subwavelength systems we can also linearise the system of equations by letting $\omega = \omega_\mathrm{lsp}$.
We can then solve,
\begin{align}\label{eqn:lin_gf}
    \left(\frac{1}{\alpha(\omega)}\mathbb{I} - \hat{\mathbf{H}}(\mathbf{k}_\parallel, \omega_
    \mathrm{lsp})\right)\cdot\mathbf{p} = 0.
\end{align}
This is a valid approximation for small NPs, where $\omega$ varies faster in the polarisability than in the Green's functions. From the analytical expression of the polarisability we retrieve the frequencies of the eigenmodes from the above equation. 

\section{Topological Valley edge states}

\subsection{Localisation of Berry curvature at valleys}

\begin{figure}
    \centering
    \includegraphics{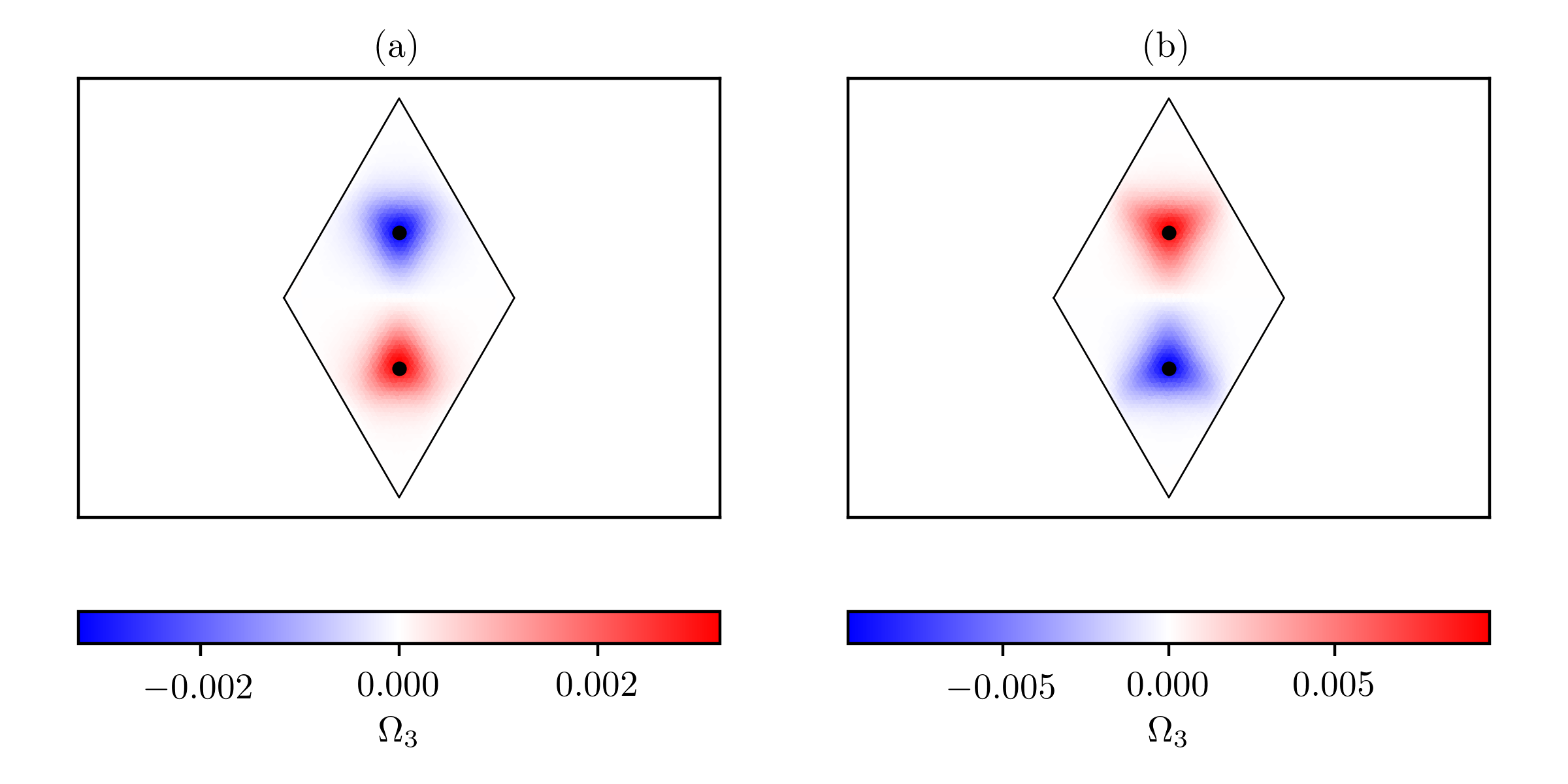}
    \caption{Localisation of Berry curvature of the highest energy band $\Omega_3$ for (a) contracted and (b) expanded lattices. The valleys ($K/K'$) are indicated as black dots.}
    \label{fig:sm_berry_curvature}
\end{figure}

The bands are calculated in a QS approximation including all neighbours and we plot these in the main text, Fig.~1(b).
We calculate the Berry curvature of the highest energy band (band 3) for the contracted and expanded lattices, $\Omega_3$.
Since this band is isolated (with no degeneracies with other bands), we use the four-point formula~\cite{depaz2020tutorial}.
For the contracted lattice in Fig.~\ref{fig:sm_berry_curvature}(a), the Berry curvature is localised around the valleys ($K/K'$), shown as black dots.
As expected for a topological valley system, the Berry curvature has opposite sign (indicated by the red/blue colormap) at opposite valleys.
The expanded lattice in Fig.~\ref{fig:sm_berry_curvature}(b) also has relatively good localisation of Berry curvature, but now the sign flips at the respective valleys compared to the contracted lattice.
This flipping of the sign between same valleys of either lattice is critical for the realisation of topological valley modes.
The Chern number $\mathcal{C}$ is the sum of Berry curvature across the Brillouin zone (BZ).
Since the system is time-reversal invariant, the total Berry curvature and so the Chern number $\mathcal{C} = 0$, as we expect.
If we split the BZ regions shown in Fig.~\ref{fig:sm_berry_curvature} in half, with the $K$ valley in the top half and $K'$ valley in the bottom half and integrate the Berry curvature then we find the valley Chern number  $\mathcal{C}_V$.
In Fig.~\ref{fig:sm_berry_curvature}(a) we have $\mathcal{C}_K = -\mathcal{C}_K' = 0.50$ as expected (and vice versa for Fig.~\ref{fig:sm_berry_curvature}(b)).

Recall that the perturbation $\delta$ controls the localisation of Berry curvature.
There is a tradeoff between having a large enough perturbation $\delta$ such that band gap which is large enough to host edge modes, and a small enough $\delta$ such that the Berry curvature is well localised.
While we could choose a smaller $\delta$ here to have better localisation, the lossy nature of the metallic system restricts how small the band gap can be to still observe edge states.

\subsection{Remarks on the ribbon calculation}

In the main text, we explained that in order to calculate edge states we set up a doubly-periodic ribbon.
To prevent extra diffraction orders entering the band gap, as a result of periodicity perpendicular to the interface, we remove large wavenumber $k$ terms from the lattice sums.
Since the lattice sums are slowly convergent, we use Ewald's method which splits the real space sum into one partially over real space and one partially over reciprocal space~\cite{linton2010lattice}.
By restricting the reciprocal space sum to along reciprocal lattice directions \textit{parallel} to the interface, we can cut-off the extra diffraction orders.

\begin{figure}
    \centering
    \includegraphics{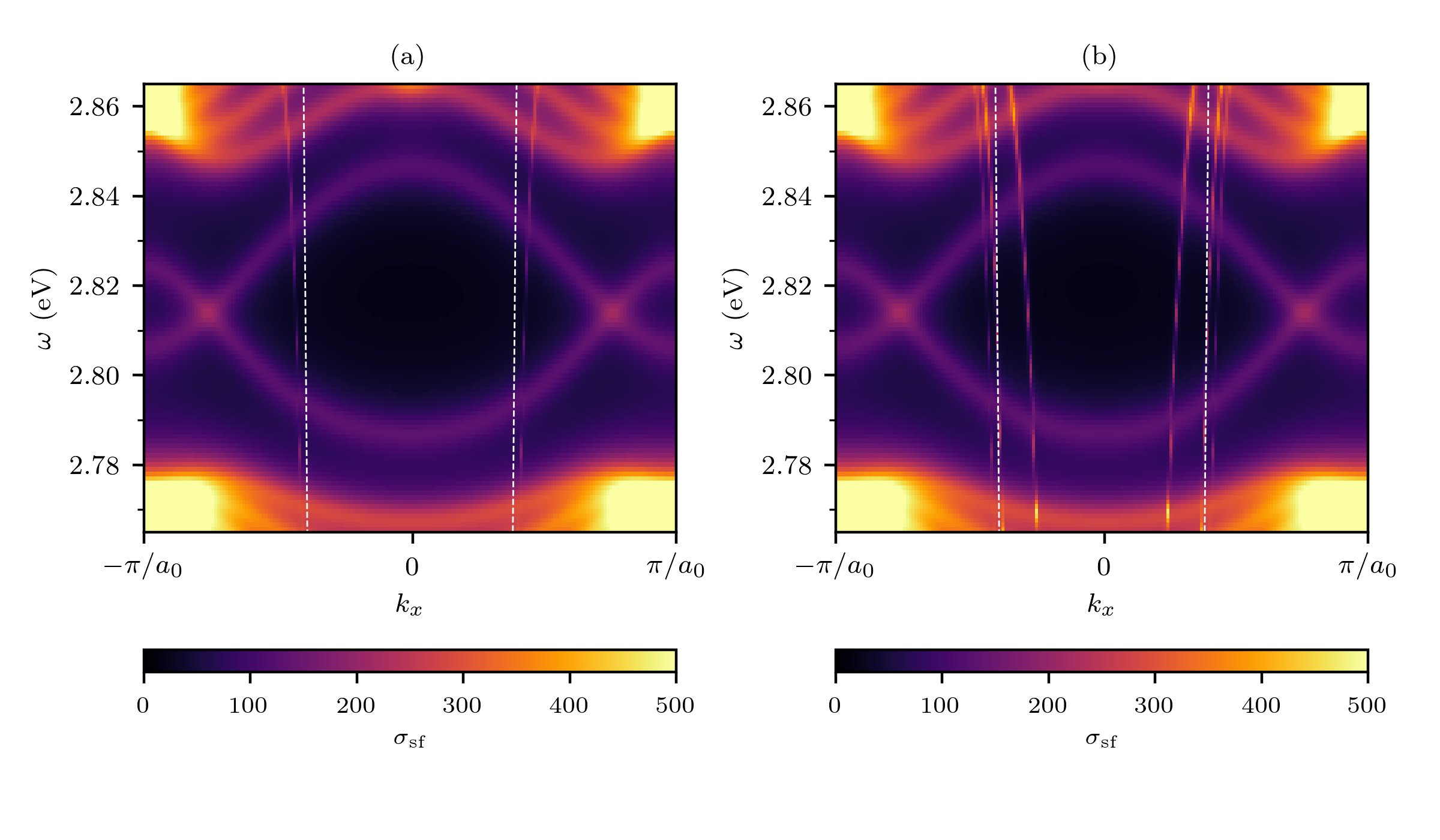}
    \caption{Spectral functions for edge modes, with the same system parameters as those presented in the main text. (a) Higher diffraction orders removed, (b) With higher diffraction orders. The edge states are unaffected by the removal of the higher diffraction orders.}
    \label{fig:sm_diffraction_orders}
\end{figure}

In Fig.~\ref{fig:sm_diffraction_orders} we show how this affects the spectral function of the edge modes.
In Fig.~\ref{fig:sm_diffraction_orders}(a), we re-plot the edge mode spectral function from the main text, using the same system parameters.
While in Fig.~\ref{fig:sm_diffraction_orders}(b), we show the spectral function when the diffraction orders are not cut-off.
We see how there are sharp features above the light line due to the extra diffraction orders.
Importantly, by cutting off the lattice sum in reciprocal space we do not affect the edge states themselves.

\subsection{Eigenmodes of the contracted over expanded interface}

\begin{figure}[h]
    \centering
    \includegraphics{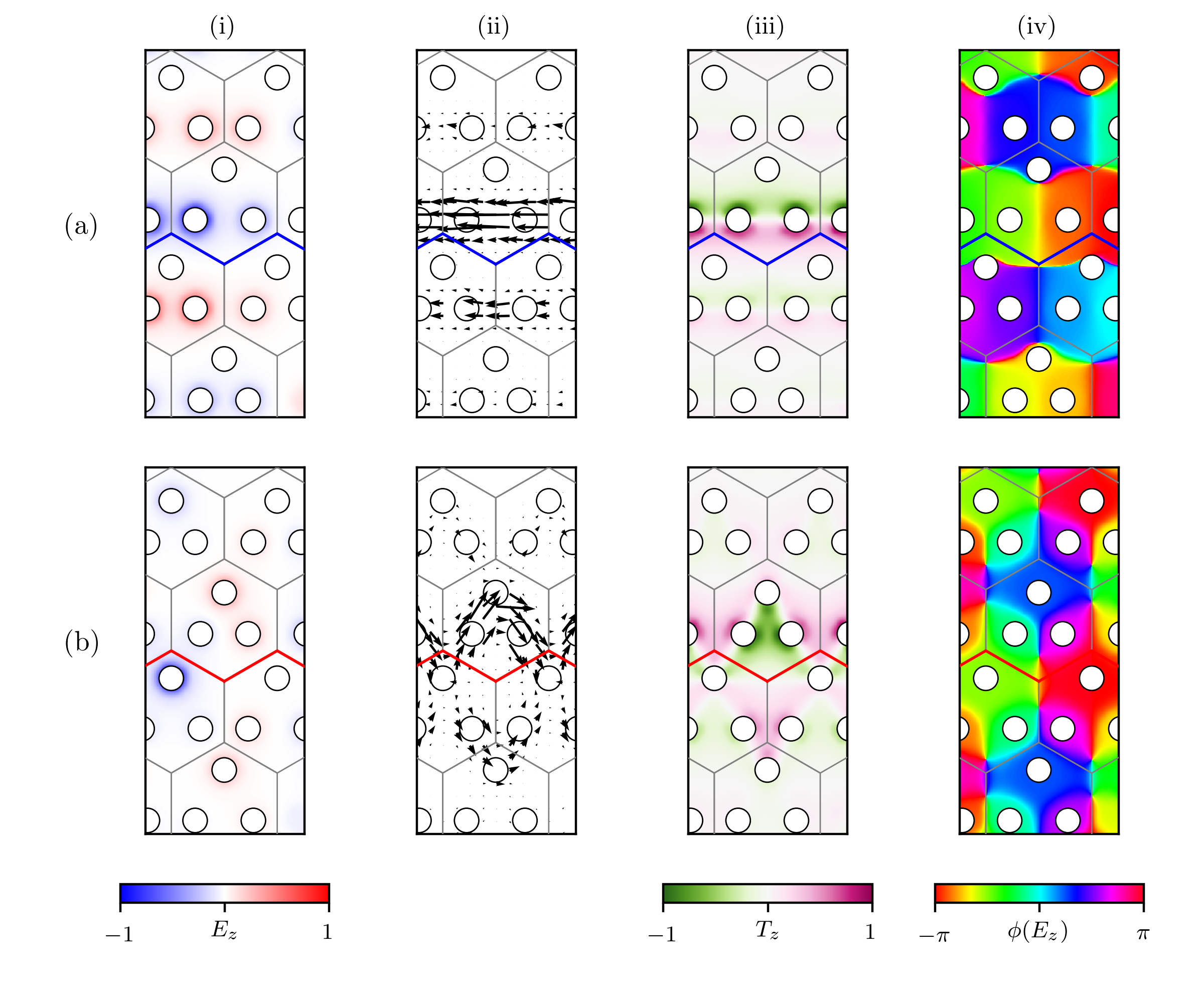}
    \caption{Edge eigenmodes of (a) expanded over contracted and (b) contracted over expanded interfaces. (i) Out-of-plane electric field $E_z$, (ii) Time-averaged Poynting vector $\langle\mathbf{S}\rangle$, (iii) Out-of-plane spin angular momentum $T_z$, and (iv) Phase of the out-of-plane electric field $\phi(E_z)$.}
    \label{fig:sm_edge_eigenmodes}
\end{figure}

The expanded over contracted (E/C) and contracted over expanded (C/E) interfaces are geometrically distinct, but both host topological valley edge states. 
In Fig.~\ref{fig:sm_edge_eigenmodes} we plot the eigenmodes for the E/C, panel (a), and C/E,  panel (b) interfaces (with the E/C also shown in the main text).
We again first plot the out of plane electric field in colum (i), then three spin-related quantities: (ii) Time-averaged Poynting vector $\langle\mathbf{S}\rangle$, (iii) Out-of-plane spin angular momentum $T_z$, and (iv) Phase of the out-of-plane electric field $\phi(E_z)$.
While the edge mode is tightly confined to the interface in both cases, as shown by the electric field, it is interesting to note the difference in energy propagation along the interfaces.

\subsection{Excitation of directional modes along interfaces}

\begin{figure}[h]
    \centering
    \includegraphics{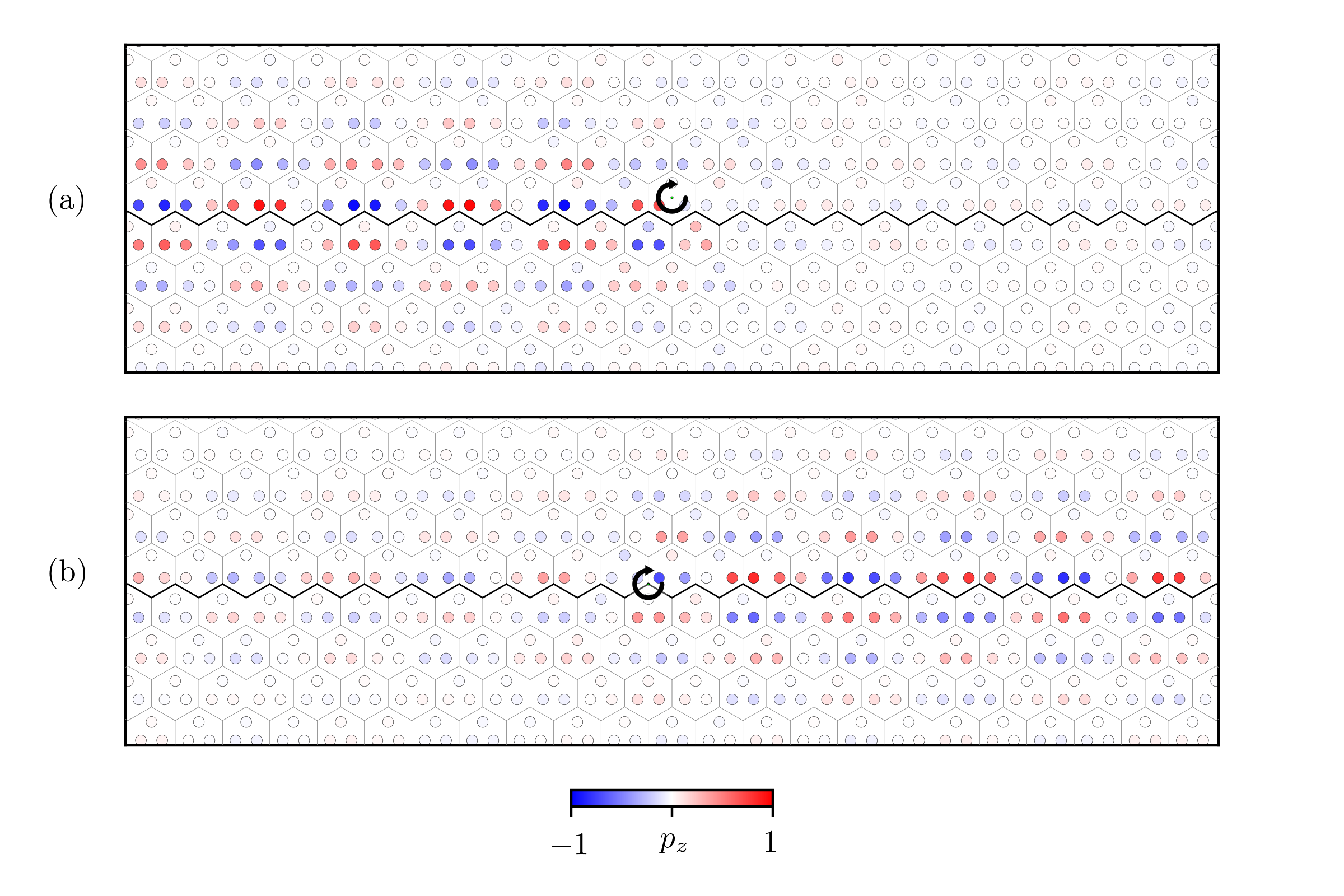}
    \caption{Directional excitation of modes with near-field sources.
    A left circularly-polarised magnetic dipole with dipole moment $\mathbf{m} = m_x - im_y$ is placed in the metasurface at,
    (a) Centre of an expanded unit cell, and
    (b) Directly at the interface. The directionality of propagation is determined by the local handedness of the spin angular momentum of the edge eigenmode.}
    \label{fig:sm_directionality}
\end{figure}

Directional propagation along interfaces between expanded and contracted regions is achieved with a circularly-polarised magnetic dipole.
In Fig.~\ref{fig:sm_directionality}, we excite with a left-circularly polarised magnetic dipole with dipole moment $\mathbf{m} = m_x - im_y$.
Firstly, in Fig.~\ref{fig:sm_directionality}(a) we place the source at the centre of an expanded unit cell directly at the interface.
In this region, the handedness of the source matches the local handedness of the magnetic field helical polarisation of the edge state eigenmode|which as we showed was determined by the spin angular momentum, $T_z$.
The matched handedness results in propagation along a direction given by the polarisation of the source, i.e. to the left.
In Fig.~\ref{fig:sm_directionality}(b), we place the source at the edge of the unit cell directly at the interface.
Now there is a handedness mismatch and so the directionality is in the opposite direction to that expected for a left-polarised source|the mode propagates to the right.

\section{Corner modes}

\subsection{Effect of increasing perturbation on corner modes}

\begin{figure}
    \centering
    \includegraphics{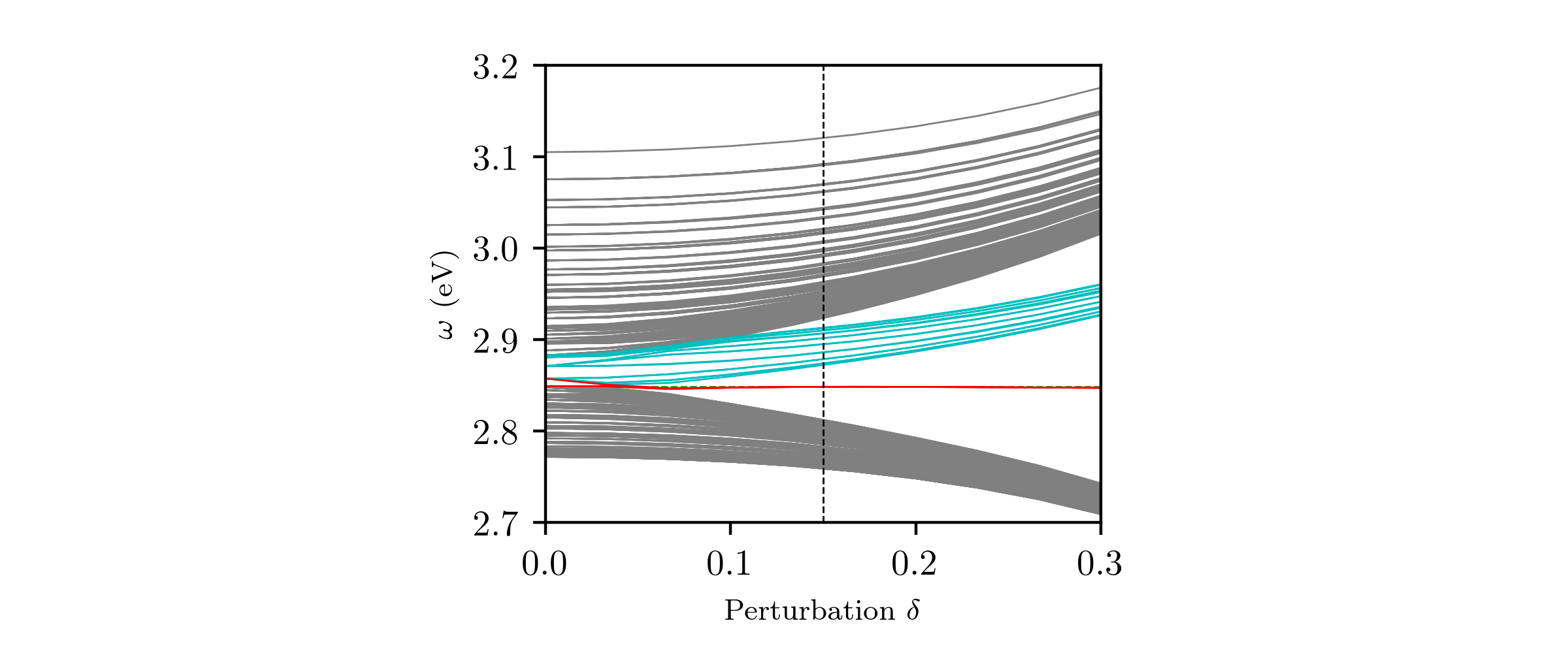}
    \caption{Frequency spectrum at normal incidence against perturbation $\delta$. Modes are coloured according to their character: Corner modes (red), edge modes (cyan) and bulk modes (grey).}
    \label{fig:sm_perturbation}
\end{figure}

While the perturbation $\delta$ controls the localisation of Berry curvature and effectiveness of edge states, in the context of HOTMs it controls the isolation of the modes in the frequency spectrum (and localisation in real space).
In Fig.~\ref{fig:sm_perturbation} we plot the frequency spectrum of the supercell at $\Gamma$ for continuously increasing perturbation.
While the bulk band gap increases in size, and the edge states move away from $\omega_\mathrm{lsp}$ (mid gap energy) the corner states are pinned to this frequency due to generalised chiral symmetry.

\subsection{Fractional corner anomaly}

Topological crystalline insulators which host corner states but do not possess symmetries such as chiral or particle-hole symmetry, will not have their corner states pinned to a mid-gap energy.
Another robust indicator of higher-order topology is the fractionalisation or splitting of charge at the corners as a result of a filling anomaly (the equal contribution of electrons to the corner modes from the valence and conduction bands)~\cite{benalcazar2019quantization}.
In a photonic system, the analogue of this is the mode density fractionalisation.
On the other hand, in higher-order systems which do not have corner states in a band gap it is not possible to probe this quantity directly.
Therefore another indicator has been proposed, a `fractional corner anomaly' (FCA), and this quantity has been measured experimentally in higher-order topological metamaterials~\cite{Peterson2020fractional}.
In an obstructed atomic limit system, the FCA is then closely related to the Wannier centres (and therefore the Wilson loops) of the bulk system~\cite{benalcazar2019quantization}.
We will show here that while the breathing kagome metasurface with photonic interactions does not have a filling anomaly or charge fractionalisation, it does possess a non-zero FCA.

\begin{figure}
    \centering
    \includegraphics{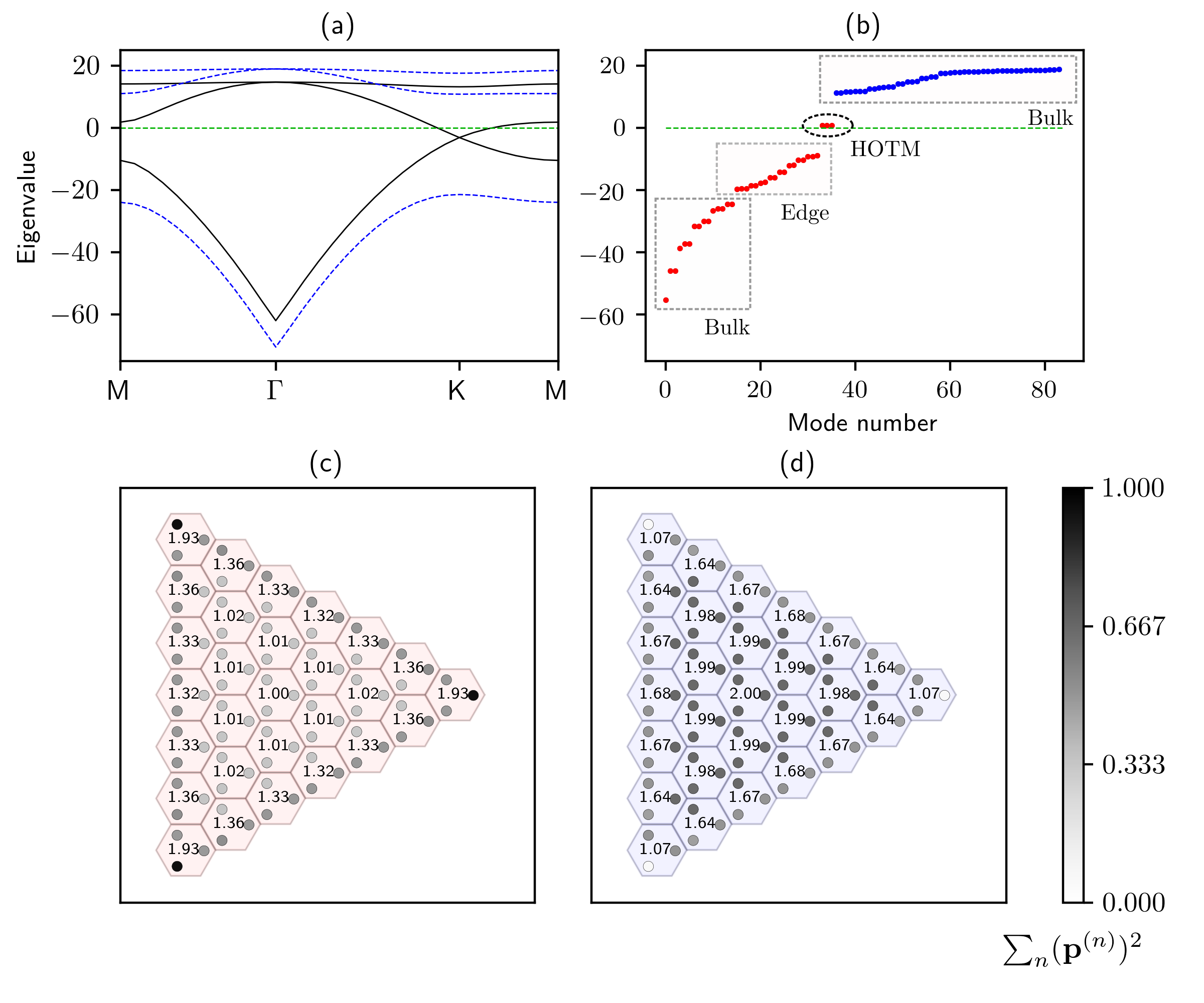}
    \caption{Fractional corner anomaly. A minimal, QS model is set up and the lattice constant is normalised to $a_0 = 1$. 
    (a) Bulk band structures for the unperturbed ($\delta = 0$, black) and perturbed ($\delta = 0.2$, dashed blue) lattice. Note that we plot eigenvalue directly. 
    (b) Eigenvalue spectrum for the topological particle.
    (c) Calculated mode densities of each unit cell for the red modes in (b) and (d) Mode densities for the blue modes in (b).}
    \label{fig:sm_fca}
\end{figure}

We set up a minimal model which incorporates long-range interactions in a quasistatic approximation.
We let the perturbation $\delta = 0.2$ and normalise the system such that the lattice constant $a_0 = 1$.
For completeness we plot the bulk band structures in Fig.~\ref{fig:sm_fca}(a), noting that they are `flipped' vertically with respect to those presented in the main text and this is because we plot eigenvalue here, rather than frequency.
The unperturbed lattice eigenvalues are plotted as solid black lines and the perturbed eigenvalues as dashed blue lines.
Next, we set up a topological particle in the expanded obstructed atomic limit phase, but unlike the main text do not include a contracted trivial `cladding' in order to simplify the model (although including this would not affect the results).
In Fig.~\ref{fig:sm_fca}(b) we plot the eigenvalue spectrum, clearly showing three HOTMs close to zero energy as well as distinct edge and bulk modes.

Next, to calculate the FCA, we first calculate the eigenmodes of all modes below zero energy plus the corner modes and all modes above zero energy (excluding the corner modes). 
These are shown as red and blues dots respectively in the eigenvalue spectrum in Fig.~\ref{fig:sm_fca}(b).
The modes given by red dots are then analogous to adding an extra `electron' to the photonic atom, or unit cell.
Immediately we see that the system does not have a filling anomaly, as there are an unequal number of modes in the bands above and below zero energy (excluding the corner modes).
The mode density $\beta$ in each unit cell is,
\begin{align}
    \beta = \sum_{i=1}^{3}\sum_n(\mathbf{p}_i^{(n)})^2.
\end{align}
The index $i$ sums over the three NPs in each unit cell and the index $n$ corresponds to the mode number.
The FCA $\phi$ is,
\begin{align}
    \phi = (\rho + 2\sigma)\mod1,
\end{align}
where $\rho$ is the mode density in the corner unit cells and $\sigma$ is the mode density in the edge unit cells (the factor of two accounts for two edges intersecting to make a corner and we assume all edges are identical).

Let us first look at the red modes: We plot the mode densities in each unit cell in Fig.~\ref{fig:sm_fca}(c).
Firstly, the bulk mode density (the unit cells within the bulk of the topological particle) is $\mu = 1$.
This is explained from the bulk eigenvalue spectrum in Fig.~\ref{fig:sm_fca}(a), where there is only one bulk band below the band gap.
Next, the edge mode density $\sigma \approx 1.33$ and the corner mode density $\rho \approx 2$.
The system does not have a mode density fractionalisation since $\rho \mod 1 = 0$.
However, the FCA is nonzero: $\phi \approx (2 + 2.66)\mod1 = \frac{2}{3}$.
For the blue modes, we plot the mode densities in Fig.~\ref{fig:sm_fca}(d).
Here the bulk mode density is $\mu = 2$ corresponding to the two bulk bands, as shown in Fig.~\ref{fig:sm_fca}(a).
The edge and corner mode densities are $\sigma \approx 1.66$ and $\rho \approx 1$, respectively, giving a FCA $\phi \approx (2 + 3.33)\mod1 = \frac{1}{3}$.

Therefore while the breathing kagome lattice does not display a fractional mode density at the corners, it does present a non-zero FCA which is evidence of the corner modes being signatures of higher-order topology.

\subsection{Exciting delocalised corner modes with near-field sources}

\begin{figure}
    \centering
    \includegraphics{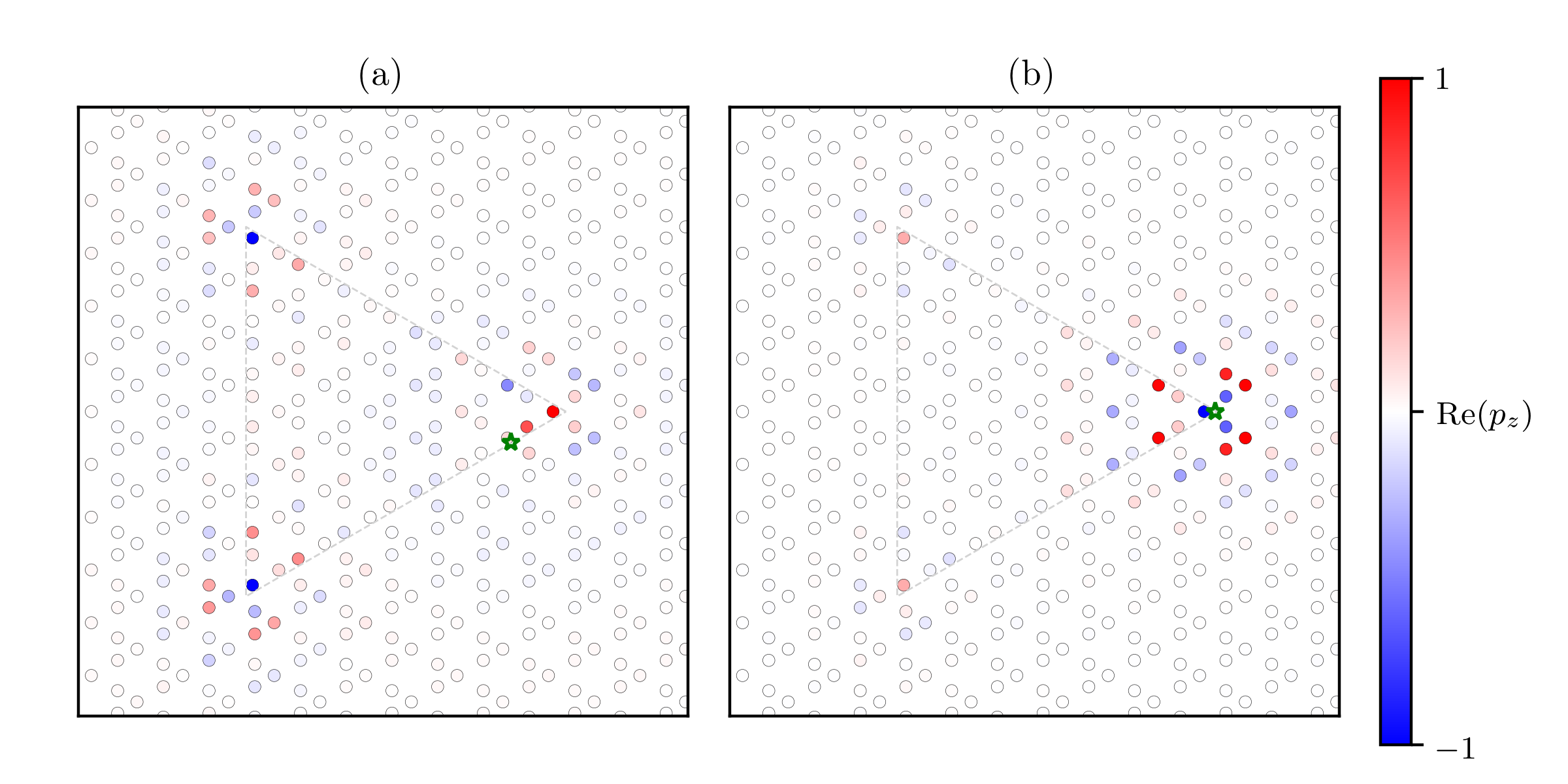}
    \caption{Exciting HOTMs with near field sources. Out-of-plane dipole moments $p_z$ for (a) source placed above a NP on the correct sublattice on the edge and (b) source placed above the NP at the corner. In each case the source is at height $z = 60nm$.}
    \label{fig:sm_delocalised}
\end{figure}

In the main text we present LDOS plots showing that at the appropriate frequency the coupling of the out-of-plane electric dipole source to the metasurface is larger when it is placed in the region around the corner. Here we show that this is indeed due to the source coupling to the corner modes, and that these corner modes are delocalised over the topological particle.
In Fig.~\ref{fig:sm_delocalised}, we place a source above the metasurface and excite at the HOTM frequency $\omega = 2.7925$~eV.
In panel (a), when the source is above the correct sublattice site close to the corner it excites the opposite corners of the topological particle `non-locally', due to coupling to the HOTM.
Similarly, when placed directly above the corner, as in panel (b), we are still able to couple to the HOTM, although the excitation is slightly weaker.
Combining this with the results in the main text, we deduce that the increased Purcell factor and LDOS when the source is exactly on top of the corner compared to the case where it is next to the corner is therefore due to coupling to the LSP in the isolated NP at corner as well as to the HOTM.

\clearpage
\bibliography{main}